\documentclass[aps,preprint,superscriptaddress,floatfix,amsmath,amssymb]{revtex4}
\usepackage{inputenc}
\usepackage[a-1b]{pdfx}
\usepackage{amsmath}
\usepackage{amssymb}
\usepackage{amsthm}
\usepackage{bm}
\usepackage{graphicx}
\usepackage{subfig}
\usepackage{hyperref}
\usepackage[capitalise]{cleveref}

\begin{document}

\title{Dynamics of a Binary Option Market with Exogenous Information and Price Sensitivity}

\author{Hannah Gampe}\email{heg5120@psu.edu}
\author{Christopher Griffin}\email{griffinch@psu.edu}
\affiliation{Applied Research Laboratory, Penn State University, University Park, PA 16802} 

\date{May 17, 2022}

\begin{abstract} In this paper, we derive and analyze a continuous of a binary option market with exogenous information. The resulting non-linear system has a discontinuous right hand side, which can be analyzed using zero-dimensional Filippov surfaces. Under general assumptions on purchasing rules, we show that when exogenous information is constant in the binary asset market, the price always converges. We then investigate market prices in the case of changing information, showing empirically that price sensitivity has a strong effect on price lag vs. information. We conclude with open questions on general $n$-ary option markets. As a by-product of the analysis, we show that these markets are equivalent to a simple recurrent neural network, helping to explain some of the predictive power associated with prediction markets, which are usually designed as $n$-ary option markets.
\end{abstract}

\maketitle

\section{Introduction}

Market models have been studied extensively both through simulation and in theory in the econophysics literature \cite{CS01,BVS06,SZSL07,S01,SP08,RP03,MZ08,CS01a,HP02,TFM18}. In contrast to traditional stock or bond markets, prediction markets have assets corresponding to future events (e.g., elections \cite{BFR97}, sports outcomes \cite{TZ88} etc.) that can be bought and sold causing changes in the underlying asset prices. Asset prices may be interpreted as probabilities \cite{M06,WZ06}. Prediction markets were first studied by Hanson \cite{H90,H91,H95,R97}. Since this initial work, they have been studied extensively \cite{WZ04,SWPG04,M06,BR03,WZ06,DJK20,CD16}. For a survey of work in this area through 2007 see \cite{TT07}. These prediction markets are archetypical examples of binary option markets (in the two outcome case) \cite{BL78}. Their use on Wall Street dates to the 1800's \cite{RS04}. Artificial prediction markets were first explored by Chen et al. \cite{CFLP08,CV10,ACW11}. This work shows a connection between learning and the prediction market showing that if a prediction market has a cost function with bounded loss, then it has an interpretation as a no-regret learning algorithm \cite{CV10}. This work is generalized in \cite{ACW11}. In the finance literature, binary options are usually used as a theoretical construct since they have a particularly simple Black-Scholes formulation \cite{H03}.

In this paper, we study the potential dynamics of prediction markets under continuum limits. In particular, we generalize and then extend the market model described in \cite{NMGR21} to a continuous time dynamic. The results of this paper are:
\begin{enumerate}
\item We show that under a constant information assumption, all asset prices converge. This explains empirical results found in \cite{NMGR21}. The result follows from an argument on Filippov surfaces. 

\item We then study the impact of dynamic information on market dynamics and empirically quantify the effect of price sensitivity and on lag between the market price and the information signal.
\end{enumerate}

%%ADD cross-references.
The remainder of the paper is organized as follows: We show the model derivation in \cref{sec:Model}. In \cref{sec:Asymptotic} we show that with constant information, all binary prediction markets converge to a constant price. In \cref{sec:Numerical}, we study the numerical results of a non-constant information. We conclude and present future directions in \cref{sec:Conclusions}.

\section{Model}\label{sec:Model}
Assume $\mathbf{q}_t = \langle{q^0(t),q^1(t)}\rangle$ units of Asset 0 and Asset 1 at time $t$ have been sold. We assume the market is composed of a collection of agents with infinite cash and that the market price is fixed by a logarithmic market scoring rule (LMSR) so that the spot prices are given by:
\begin{align}
&p^0(t) = \frac{\exp{[\beta q^0(t)]}}{\exp{[\beta q^0(t)]} + \exp{[\beta q^1(t)]}}\label{eqn:dotp0}\\
&p^1(t) = \frac{\exp{[\beta q^1(t)]}}{\exp{[\beta q^0(t)]} + \exp{[\beta q^1(t)]}}.
\label{eqn:dotp1}
\end{align}
Notice because a Boltzmann distribution is being used we have $p^0(t) + p^1(t) = 1$ for all time. Here, $\beta$ term is a liquidity factor \cite{LS18} that adjusts the amount the price will increase or decrease given a change in the asset quantities. We denote the spot-price vector $\mathbf{p}(t) = \langle{p^0(t),p^1(t)}\rangle$. Assume that $\mathbf{p}(0) = \langle{p^0_0,p^1_0}\rangle$.
 
In general, trade costs are not given by $\mathbf{p}_t$, since LMSR incorporates a market maker cost. The trade costs are given by:
\begin{gather*}
\kappa_t^0(\Delta q^0) = \frac{1}{\beta} \log\left\{\frac{\exp[\beta (q^0(t)+\Delta q^0)] + \exp[\beta q^1(t)]}{\exp[\beta q^0(t)] + \exp[\beta q^1(t)]}\right\}\\
\kappa_t^1(\Delta q^1) = \frac{1}{\beta} \log\left\{\frac{\exp[\beta q^0(t)] + \exp[\beta (q^1(t) + \Delta q^1)]}{\exp[\beta q^1(t)] + \exp[\beta q^0(t)]}\right\},
\end{gather*}
where $\Delta q^i$ is the change in the quantify of Asset $i$ as a result of purchases by an agent. For mathematical simplicity we will assume agents purchase one share at a time so that $\Delta q^i \in \{0,1\}$. Let:
\begin{equation*}
\kappa_t^i = \kappa_t^i(1)
\end{equation*}
be the price of single share purchase of Asset $i$. 

\subsection{Agent Purchase Logic and Information}
Following \cite{NMGR21}, we assume agents specialize in the purchase of a specific asset class (either $0$ or $1$) and that there is some (possibly time-varying) external information $\mathbf{x}(t) \in \mathbb{R}^n$ that will inform this purchase decision. 

Assume agents are indexed in $\{1,\dots,N\}$ and suppose that agent $i$ purchases only assets in class $y_i \in \{0,1\}$. We assume agents have an infinite cash reserve and do not sell assets back to the market maker. Following \cite{NMGR21}, we assume that each agent uses a characteristic function $\psi_i:\mathbb{R}^n \times \mathbb{R}^2 \to \mathbb{R}$ to reason about information $\mathbf{x}$ so that its decision to buy an asset in class $y_i$ is governed by:
\begin{equation}
\Delta q^{y_i}_i = H\left\{\sigma[\psi_i(\mathbf{x},\mathbf{p})] - \kappa^{y_i}\right\}.
\label{eqn:PurchaseEqn}
\end{equation}
Here $\sigma:\mathbb{R}\to[0,1]$. In \cite{NMGR21}, this is assumed to be a logistic sigmoid function:
\begin{equation*}
\sigma(x) = \frac{1}{1+e^{-x}},
\end{equation*}
and $H(x)$ is the unit step function defined as $0$ at $x = 0$. We note that any mapping $\sigma:\mathbb{R}\to[0,1]$ can be used in the remainder of this paper. The expression $\sigma[\psi_i(\mathbf{x},\mathbf{p})]$ defines the value Agent $i$ places on Asset $y_i$ as a function of the current asset price(s) and the information $\mathbf{x}$. Thus $\Delta q^{y_i}_i = 1$ just in case:
\begin{equation*}
\sigma[\psi_i(\mathbf{x},\mathbf{p})] > \kappa^{y_i}.
\end{equation*}
\cite{NMGR21} assumes that $\mathbf{x}$ is a vector in an appropriate feature space and for some scalar $R$ proposes that $\psi(\mathbf{x},\mathbf{p}) \geq R$ should define a convex region in $\mathbb{R}^n \times \mathbb{R}^2$. Thus, a collection of (price sensitive) agents encode a set of time-varying sets in feature space and an agent purchases a share of a relevant asset if and only if the information encoded in the (constant) feature and price vector $(\mathbf{x},\mathbf{p})$ is an element of the region defined by $\sigma[\psi(\mathbf{x},\mathbf{p})]$. \cite{NMGR21} shows that price insensitive agents can be used to define an arbitrary covering region to encode a binary classification problem but observes in discrete time that minor oscillations can occur in the price. We show that this cannot happen in continuous time assuming a constant feature vector and then explore the affects of a non-constant feature vector.

\subsection{Continuous Model}

Consider $\kappa^1_t(\Delta q^1)$. We can rewrite this as:
\begin{multline*}
\kappa_t^1(\Delta q^1) = \frac{1}{\beta} \log\left\{\frac{\exp[\beta q^0(t)] + \exp[\beta (q^1(t) + \Delta q^1(t))]}{\exp[\beta q^1(t)] + \exp[\beta q^0(t)]}\right\} = \\
\frac{1}{\beta}\log\left\{\exp\left({\beta\Delta q^1(t)}\right)\frac{\exp[\beta q^1(t)]}{\exp[\beta q^1(t)] + \exp[\beta q^0(t)]} + \frac{\exp[\beta q^0(t)]}{\exp[\beta q^1(t)] + \exp[\beta q^0(t)]}\right\} = \\
\frac{1}{\beta}\log\left[\exp\left[{\beta\Delta q^1(t)}\right]p^1(t) + p^0(t)\right].
\end{multline*}
We know that $p^0(t) + p^1(t) = 1$, so we have:
\begin{displaymath}
\kappa_t^1(\Delta q^1) = \frac{1}{\beta}\log\left\{[\exp\left({\beta\Delta q^1}\right)-1]p^1 + 1\right\}.
\end{displaymath}
For $\beta \ll 1$, we can approximate the exponential and logarithmic functions with a Taylor series: $e^x \approx 1 + x + O(x^2)$ and $\log(1+x) \approx x$ to see:
\begin{displaymath}
\kappa_t^1(\Delta q^1) = \frac{1}{\beta}\log\left[{\beta\Delta q^1}p^1(t) + 1\right] = \frac{1}{\beta}\left[\beta\Delta q^1p^1(t)\right] =\Delta q^1p^1(t). 
\end{displaymath}
Since we assume $\Delta q^1 \in \{0,1\}$, we may assume $\kappa_t^1 \approx p^1(t)$.

Following \cite{NMGR21}, let $C_1 \subset \{1,\dots,N\}$ be the set of agents that buy only Asset $1$ and let $C_0 \subset \{1,\dots,N\}$ be the set of agents that buy only Assert $0$. Using our approximation from \cref{eqn:PurchaseEqn} we have:
\begin{align*}
q^1(t+1) &\approx q^1(t) + \sum_{i \in C_1} H\left\{\sigma[\psi_i(\mathbf{x},\mathbf{p})] - p^1(t)\right\}\\
q^0(t+1) &\approx q^0(t) + \sum_{i \in C_1}H\left\{\sigma[\psi_i(\mathbf{x},\mathbf{p})] - p^0(t)]\right\},
\end{align*}
If we assume that transaction times are proportional to a time slice so that:
\begin{equation}
q^1(t+\epsilon) \approx q^1(t) + \epsilon\sum_{i \in C_1} H\left\{\sigma[\psi_i(\mathbf{x},\mathbf{p})] - p^1_t]\right\},
\label{eqn:EulerStep}
\end{equation}
then, taking the limit as $\epsilon \to 0$, we obtain the non-linear differential equation:
\begin{align}
\dot{q}^1 &= \sum_{i \in C_1} H\left\{\sigma[\psi_i(\mathbf{x},\mathbf{p})] - p^1_t]\right\}\label{eqn:ode1}\\
\dot{q}^0 &= \sum_{i \in C_0} H\left\{\sigma[\psi_i(\mathbf{x},\mathbf{p})] - p^0_t]\right\}\label{eqn:ode0}
\end{align}
Using \cref{eqn:dotp0,eqn:dotp1} and the quotient rule we can compute:
\begin{align*}
&\dot{p}^1 = \beta p^1 p^0 \left(\dot{q}^1 - \dot{q}^0\right)\\
&\dot{p}^0 = \beta p^0 p^1 \left(\dot{q}^0 - \dot{q}^1\right).
\end{align*}
These results are consistent with \cite{TVL03,BCFG22} for multi-agent systems using a Boltzmann (soft-max) learning rule assuming a two-class outcome. 

\begin{figure}[htbp]
\centering
\includegraphics[width=0.6\textwidth]{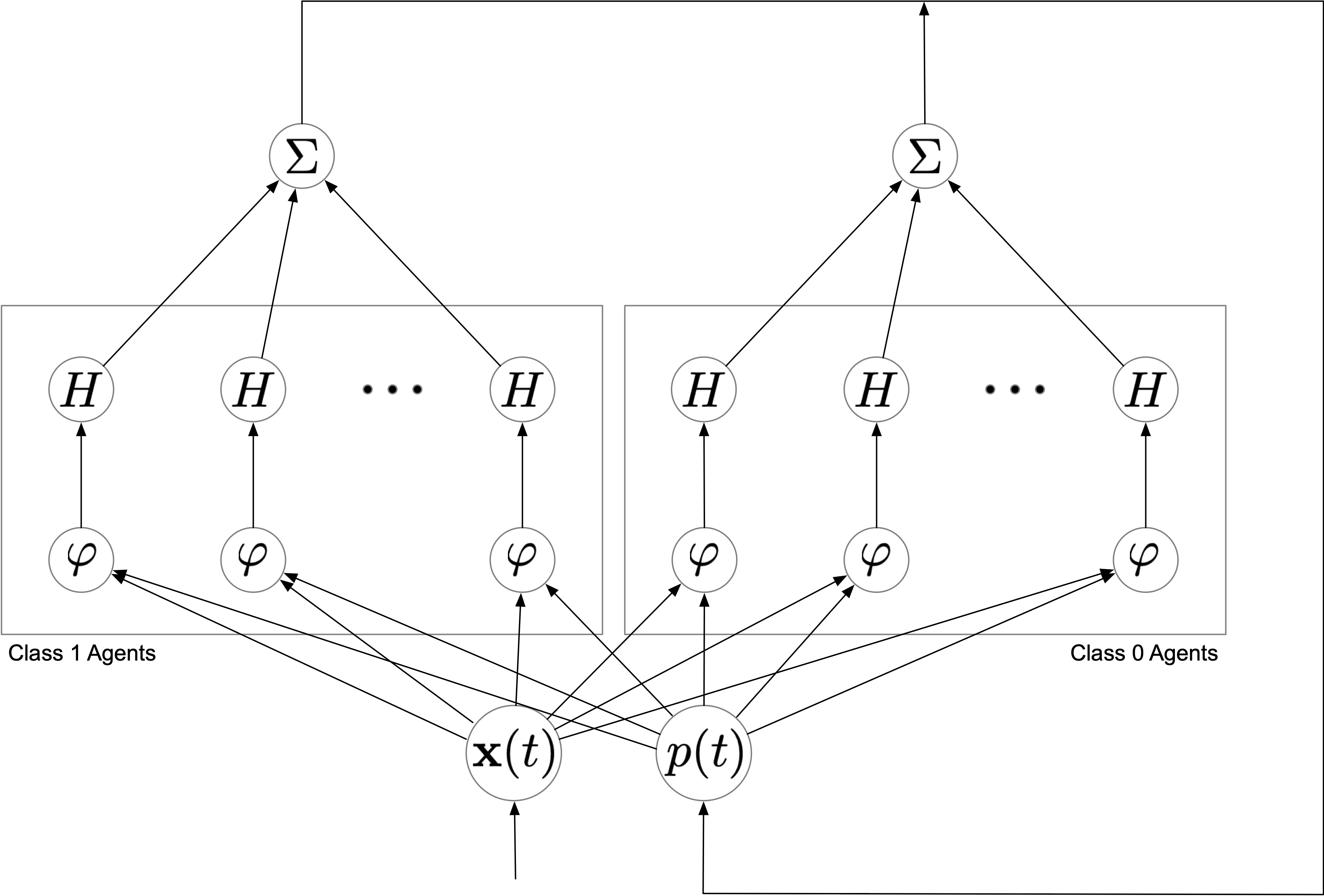}
\caption{}
\label{fig:NN}
\end{figure}

Using the fact that $p^0(t) + p^1(t) = 1$ for all time and substituting in our expressions for $\dot{q}^1$ and $\dot{q}^0$, we obtain a single ODE in terms of $p^1(t)$, which we now write simply as $p(t)$:
\begin{equation}
\dot{p} = \beta p(1-p)\left(\sum_{i \in C_1} H\left\{\sigma\left[\psi_i(\mathbf{x},p)\right] - p\right\} - \sum_{i \in C_0} H\left\{\sigma\left[\psi_i(\mathbf{x},p)\right] - (1-p)\right\} \right).
\label{eqn:ODE}
\end{equation}
Notice unlike the work in \cite{TVL03,BCFG22}, this ODE has a piecewise smooth and bounded right-hand-side with jump discontinuities. Therefore it is a Filippov system \cite{F67,F13}. Moreover, this system can be described by a continuous time recurrent neural network, shown in \cref{fig:NN}. 
The step function $H$ acts as an activation function for the non-linear neurons computing:
\begin{equation*}\varphi_i(\mathbf{x},p) = \sigma\left[\psi_i(\mathbf{x},p)\right] - p. 
\end{equation*}
Thus, the dynamics of the neural network can be completely determined the exogenously specified dynamics of $\mathbf{x}(t)$ and \cref{eqn:ODE}.

\section{Asymptotic Market Behavior for Constant Information}\label{sec:Asymptotic}
Assume $\mathbf{x}(t)$ is constant so that $\psi_i[\mathbf{x}(t)] = L_i + f_i(p)$ where $L_i \in \mathbb{R}$. The dynamics of $p(t)$ are then:
\begin{equation}
\dot{p} = \beta p(1-p)\left(\sum_{i \in C_1} H\left\{\sigma\left[L_i + f_i(p)\right] - p\right\} - \sum_{i \in C_0} H\left\{\sigma\left[L_i +  f_i(p)\right] - (1-p)\right\} \right).
\label{eqn:ODE2}
\end{equation}
The right-hand-side is piecewise smooth and bounded and therefore a solution to this ODE exists \cite{F13}. It is clear that $p^* = 0$ and $p^* = 1$ are fixed points of \cref{eqn:ODE2}. Therefore, for $p_0 \in [0,1]$, if \cref{eqn:ODE} if $p(t)$ is a solution, then $p(t) \in [0,1]$ for all time.

We have just shown that the dynamics of $p(t)$ are contained entirely in the smooth manifold $M = [0,1]$. Thus the flow at any point $M$ is defined by a scalar (since the dimension of $M$ is 1). By convention positive flow moves toward $p^* = 1$. A \textit{sliding mode} fixed point or fixed point on a Filippov surface is a point $p^* \in \mathrm{int}(M)$ (a zero dimensional manifold) where the flow to the left of $p^*$ is positive and the flow to the right of $p^*$ is negative and the flow at $p^*$ may take either a positive or negative or zero value \cite{F13}. That is, the flow around this point is moving in opposite directions. Naturally, this definition can be extended to higher dimensional flows (see \cite{F13}) but we do not require this level of complexity.

Let:
\begin{equation*}
N(p) = \sum_{i \in C_1} H\left\{\sigma\left[L_i + f_i(p)\right] - p\right\} -
\sum_{i \in C_0} H\left\{\sigma\left[L_i + f_i(p)\right] - (1-p)\right\}.
\end{equation*}
This is a discontinuous integer valued function with maximum value $|C_1|$ and minimum value $-|C_0|$. We can re-write \cref{eqn:ODE} as:
\begin{displaymath}
\dot{p} = \beta N(p) p (1-p).
\end{displaymath}
From this, we see that for any fixed integer $N(p)$, \cref{eqn:ODE} is just a logistic differential equation, which has a global solution. We deduce at once that \cref{eqn:ODE} has a global continuous solution composed of piecewise smooth solutions to the individual logistic differential equations.  

We now show that these solutions are monotonic and either asymptotically approach a value as time goes to infinity or reach a fixed point in finite time. 

\begin{figure}[htbp]
\centering
\includegraphics[width=0.45\textwidth]{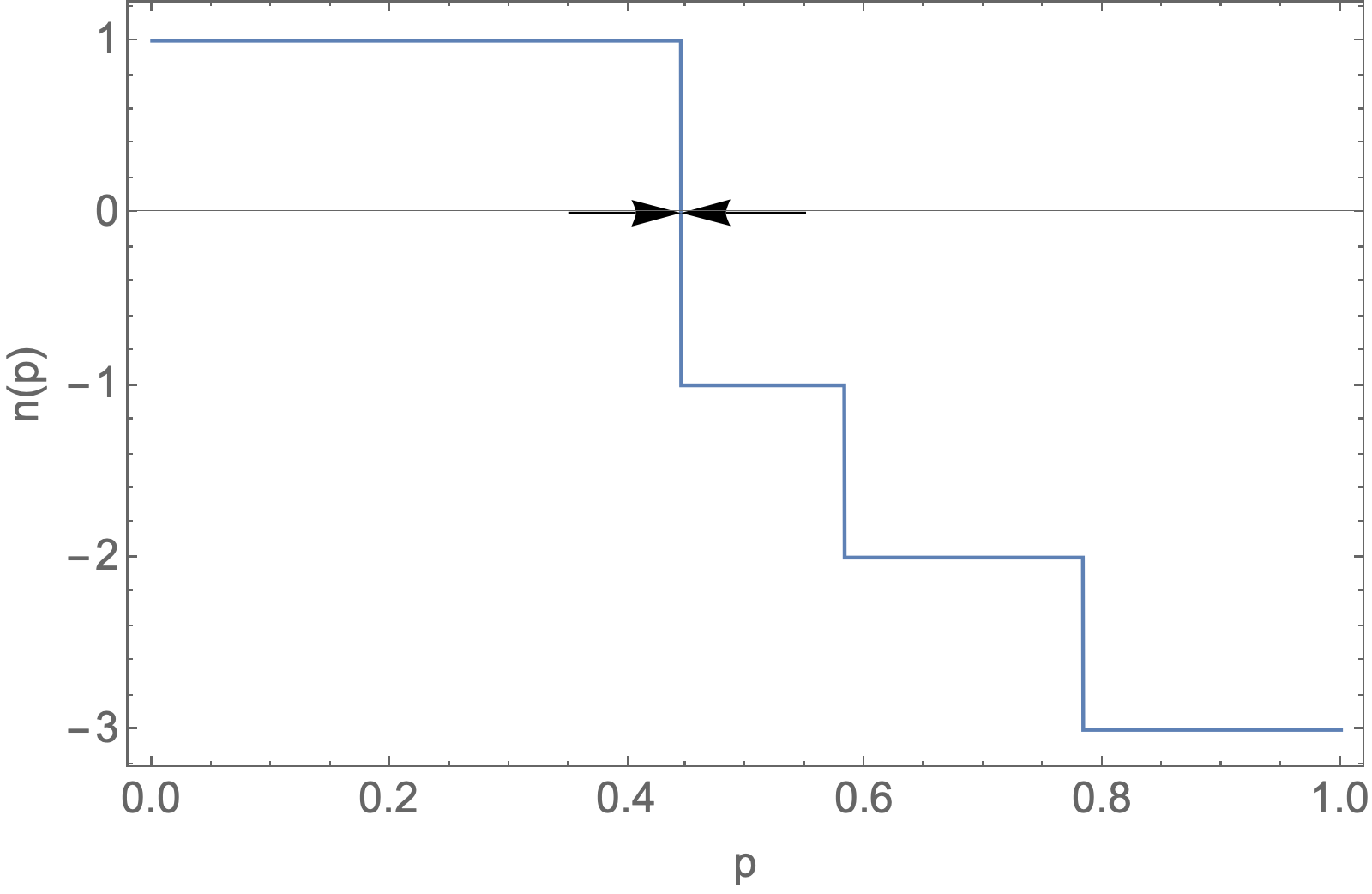}\quad
\includegraphics[width=0.46\textwidth]{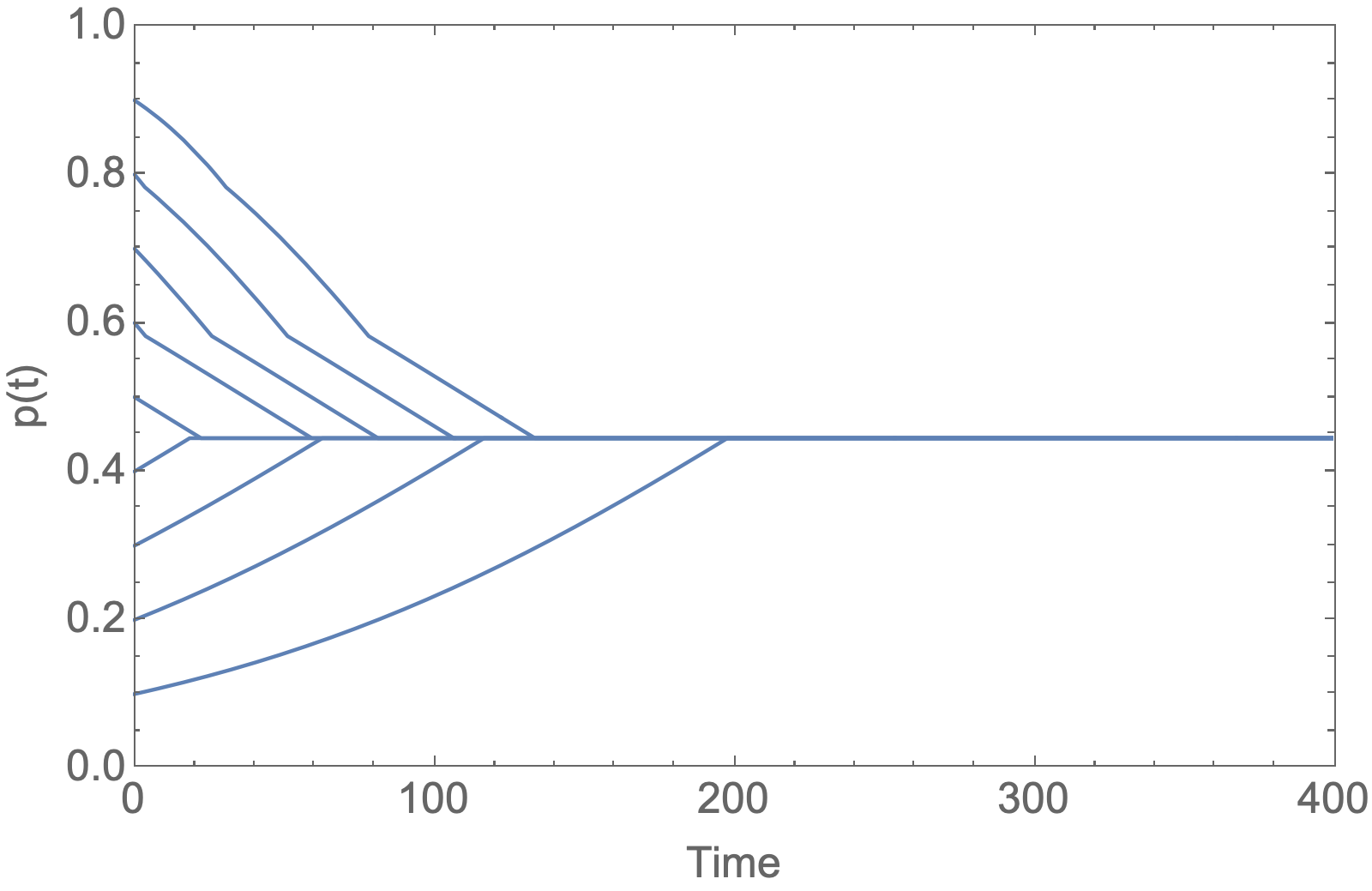}
%\textbf{Hannah, please add figures here.}
%\includegraphics[width=0.4\textwidth]{Figures/myfigure.pdf}
%\includegraphics[width=0.5\textwidth]{Figures/figure0.pdf}
%\caption{Illustration of discontinuity at zero, resulting in a Filippov sliding mode (left) and leading to zero (right).}
%\includegraphics[width=0.4\textwidth]{Figures/badarrows1.pdf}
%\includegraphics[width=0.4\textwidth]{Figures/badarrows.pdf}
\caption{Illustration of a function $n(p)$ with a jump discontinuity and the resulting convergence to the sliding mode manifold.}
\label{fig:CaseIllustration}
\end{figure}

If $N(p_0)=0$, then $p=p_0$ is the solution for all time. On the other hand, suppose $N(p_0) < 0$, then $p(t)$ is initially decreasing and we have three possibilities.
\begin{description}
\item[Case I:]  If $N[p(t)]$ remains negative as $p(t)$ decreases, then $p(t)$ decreases for all time, asymptotically approaching the fixed point $p^* = 0$. 
    
\item[Case II:] If $N[p(t)]$ increases as $p(t)$ decreases and there is a finite $t^* > 0$ such that  $N[p(t^*)] = 0$, then for all $t \geq t^*$, $p(t) = p(t^*)$. 
    
\item[Case III:] If $N[p(t)]$ increases as $p(t)$ decreases and there is a smallest time $t^* > 0$ so that for all $\epsilon > 0$, $N[p(t^*+\epsilon)] > 0$ and $N[p(t^*-\epsilon)] < 0$, then $p(t^*)$ is a sliding mode fixed point. Consequently, the value of $p(t)$ becomes fixed at the location of the jump discontinuity in $N(p)$. This is illustrated in \cref{fig:CaseIllustration}.
\end{description}
The argument for the case when $N(p_0) > 0$ is symmetric. Thus, we have shown that: (i) Every instance of \cref{eqn:ODE} has a global continuous and piecewise smooth solution. (ii) Solutions are monotonic. (iii) Every solution either asymptotically approach $p^* = 0$ or $p^* = 1$ or reaches a constant value infinite time.

\section{Numerical Results with Non-Constant Information}\label{sec:Numerical}
When $\mathbf{x}(t)$ is non-constant, the market will track the function $\psi(\mathbf{x},p)$ through the function:
\begin{equation*}
N(\mathbf{x},p) = \left(\sum_{i \in C_1} H\left\{\sigma\left[\psi_i(\mathbf{x},p)\right] - p\right\} - \sum_{i \in C_0} H\left\{\sigma\left[\psi_i(\mathbf{x},p)\right] - (1-p)\right\} \right)
\end{equation*}
If there is some time $T \geq 0$ so that for all $t \geq T$ we have $N[\mathbf{x}(t),p(t)] \geq 0$ ($N[\mathbf{x}(t),p(t)] \leq 0$, respectively) then clearly $p(t)$ must converge. 

\begin{figure}[htbp]
\centering
\includegraphics[width=0.8\textwidth]{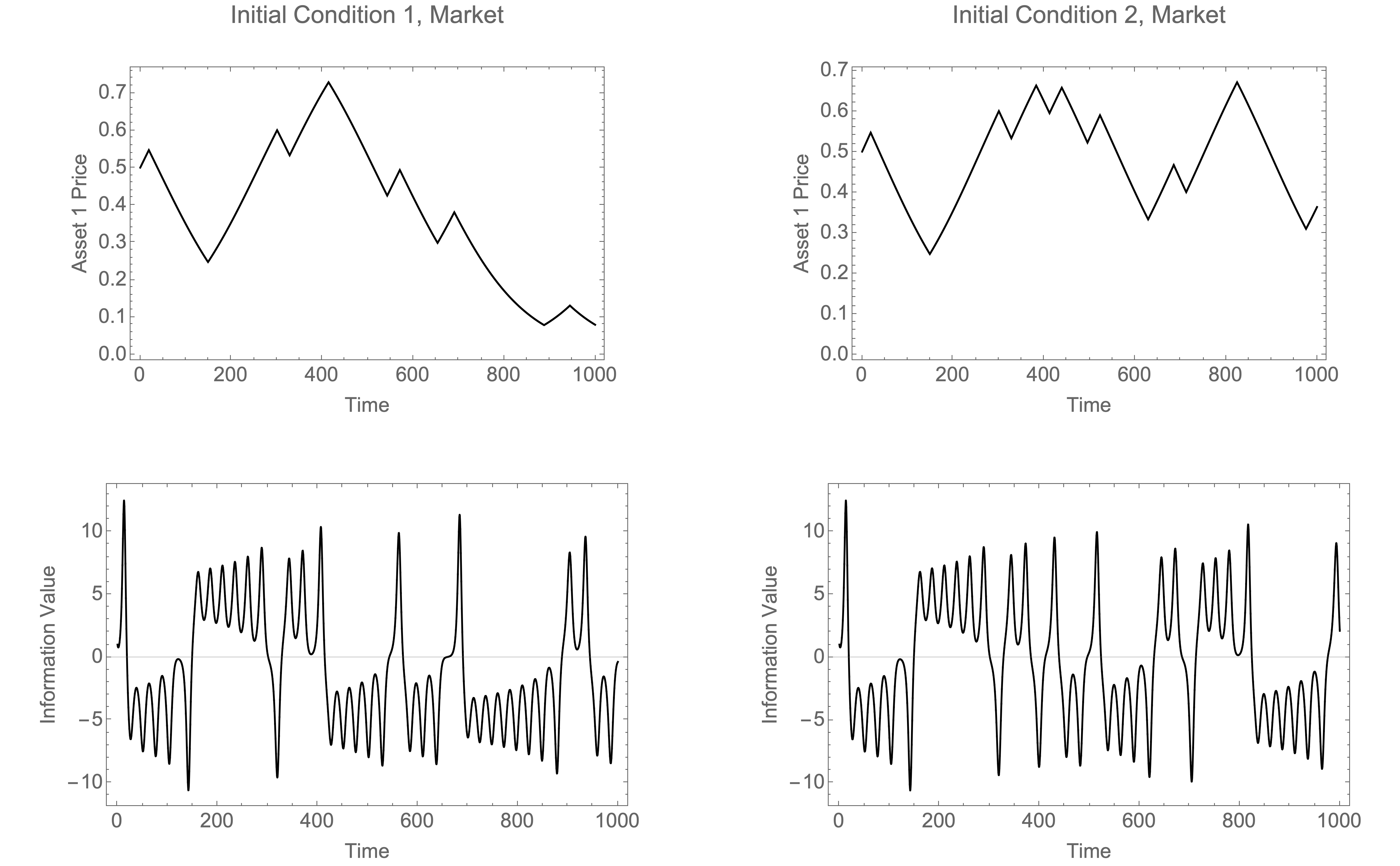}
\caption{Two separate Lorenz solutions (first dimension only) are shown one with initial condition $(1,-1,1)$ and $(1.01,-1,1)$. The market smooths the dynamics of the Lorenz system. Here, $\beta = 0.01$.}
\label{fig:ChaoticInformation}
\end{figure}

When $N(\mathbf{x},p)$ remains above (or below) $0$, the market will lag the information content in the $\mathbf{x}(t)$ and the market will smooth the information in the signal $\mathbf{x}(t)$. By way of example, let $\mathbf{x}(t)$ be a solution to the Lorenz system and define:
\begin{align*}
&\psi_1(\mathbf{x},p) = \mathbf{x}_1(t)\\
&\psi_0(\mathbf{x},p) = -\mathbf{x}_1(t),
\end{align*}
where $\mathbf{x}_1$ is the first dimension of the three dimensional curve $\mathbf{x}(t)$. Then:
\begin{equation*}
\dot{p} = \beta p (1-p)\left(H\left\{\sigma\left[\psi_1(\mathbf{x},p)\right] - p\right\} -  H\left\{\sigma\left[\psi_0(\mathbf{x},p)\right] - (1-p)\right\}\right)
\end{equation*}
The dynamics are shown in \cref{fig:ChaoticInformation} for two different initial conditions.
This shows that the market preserves the sensitive dependence on initial conditions (as expected) and smooths the dynamics of the Lorenz system. Numerical solution of the differential equation systems is described in \ref{sec:NumericalSoln}.

\begin{figure}[htbp]
\centering
\includegraphics[width=0.45\textwidth]{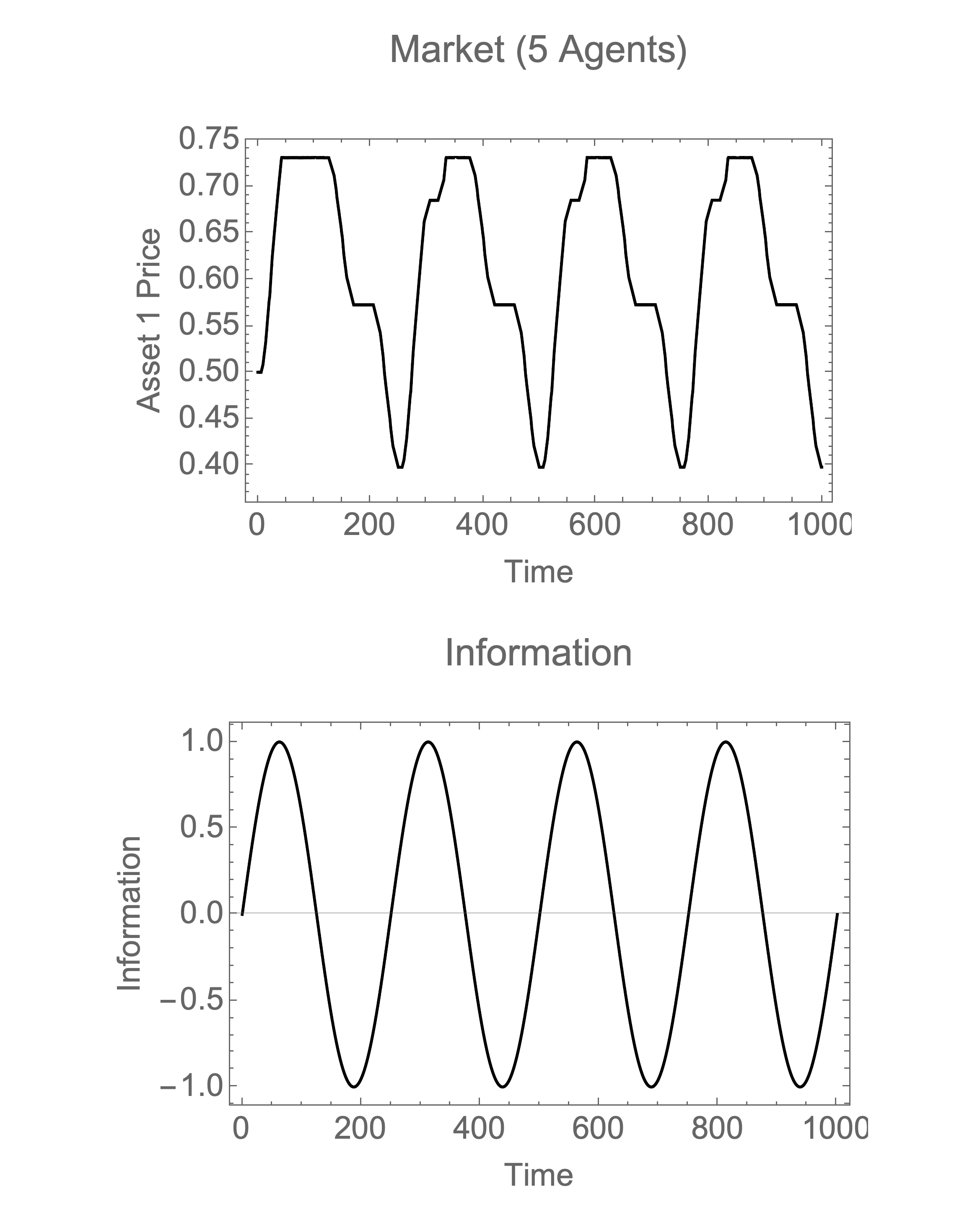}\qquad
\includegraphics[width=0.45\textwidth]{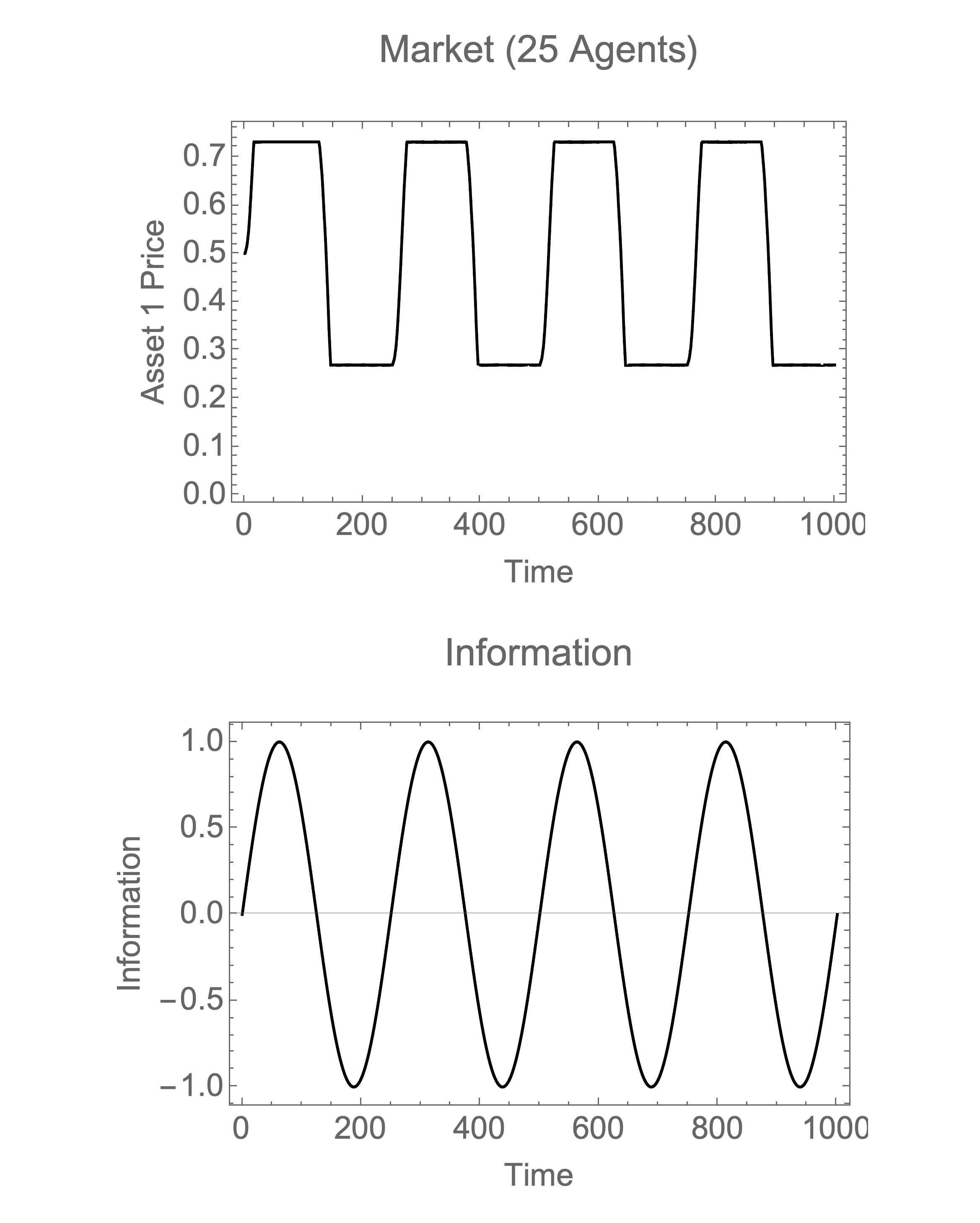}\qquad
\caption{(Left) A market with ten agents (five agents per class) shows tracking of the information $x(t)$. (Right) A market with fifty agents (25 agents per class) tracks that information $x(t)$ but the market price shows saturation.}
\label{fig:Agents}
\end{figure}

There is a lag in the market price as new information is assimilated into the asset price. To study this lag and the effect of various market conditions, we assume that $x = A \sin(\omega t)$. For our experiments, we set $A = 1$ and $\omega = 2\pi/25000$. Define a set of agents by random intervals $\{(a_i,b_i)\}_{i=1}^N$ where if $i \in C_1$, then $a_i \geq 0$ and $b_i \leq 1$ and if $i \in C_0$, then $b_i \leq 0$ and $a_i \geq -1$. When $N$ is large, with high probability these intervals cover the $[-1,1]$. Assume:
\begin{equation*}
\psi_i(x) = \begin{cases} 1 & \text{if $a_i < x < b_i$}\\
-1 & \text{otherwise}
\end{cases}.
\end{equation*}
Then the market will classify $x(t)$ as it moves from $+1$ to $-1$ over time. This is illustrated in \cref{fig:Agents} for two different agent sizes.
The market with smaller numbers of agents tracks the information but there are times when $x(t)$ is not in any interval defining an agent and so buying stops. This leads to plateaus in the data. 

\begin{figure}[htbp]
\centering
\includegraphics[width=0.75\textwidth]{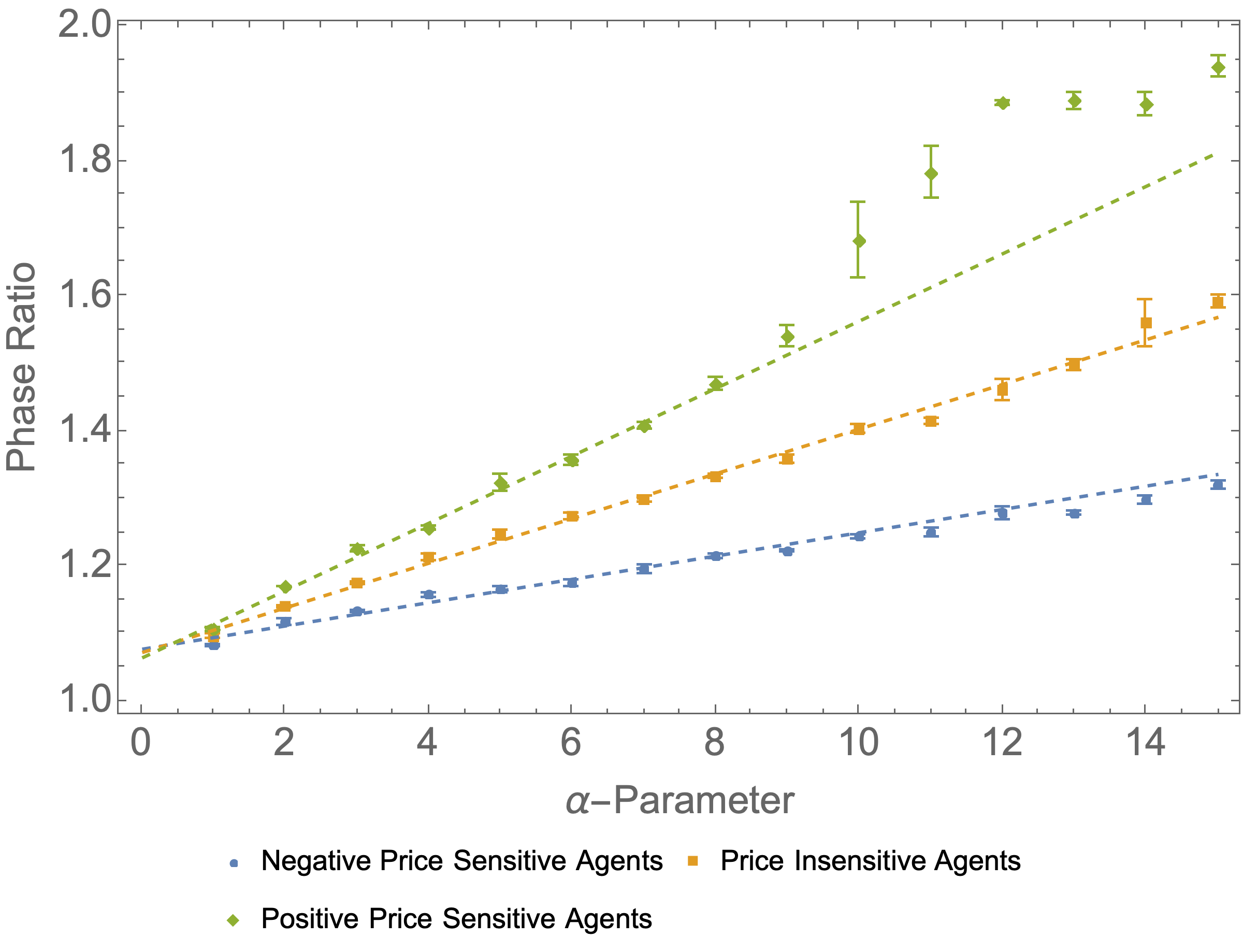}
\caption{The ratio between the phase angle at the fundamental frequency for the price and information dynamics are shown. The Line of fit it computed from the first five data points to show non-linear growth.}
\label{fig:Lag}
\end{figure}

Introduce two parameters $\alpha\geq 1$ and $\nu \in \{-1,0,1\}$ so that:
\begin{multline*}
\dot{p} = \beta p (1-p)\left(H\left\{\sigma\left[\alpha\left(\psi_1(x) + \nu\left(p-\tfrac{1}{2}\right)\right)\right] - p\right\} \right.-\\\left.  H\left\{\sigma\left[\alpha\left(\psi_0(x)+\nu\left(\tfrac{1}{2}-p\right)\right)\right] - (1-p)\right\}\right)
\end{multline*}
we can examine the effect asset price sensitivity ($\nu$) and input gain ($\alpha$) have on the lag in information uptake in the market. To do this, we use the measured phase difference between $p(t)$ and $x(t)$ using the argument of the Fourier transform of the solutions at the fundamental frequency $\omega$. The ratio of the two phases provides a measure of the speed with which information is taken up by the market. We used markets with 250 agents (125 agents per class) and allows $\alpha$ to vary from $1$ to $15$ and $\nu$ to vary in the set $\{-1,0,1\}$. For all experiments $\beta = 0.01$. When $\nu = 0$, there is no price sensitivity. When $\nu = -1$, agents are more sensitive to price and higher prices make them value the asset lower. Effectively, this creates additional friction in purchasing. When $\nu =1$, agents are price sensitive and higher prices causes them to value an asset more. This creates additional force in the the market to purchase even at higher prices. We used 5 replications of each condition to construct confidence intervals on the mean of the phase ratios. Agent intervals were randomized in each replication. Throughout all experiments, $x(t)$ was held constant with $A = 1$ and $\omega = 2\pi/25000$. Results are shown in \cref{fig:Lag}.
We conclude that agents with positive price sensitivity resulted in a higher phase ratio while agents with a negative price sensitivity had a lesser phase ratio for varying input gain ($\alpha$). Price insensitive agents' phase ratio fell between the positive and negative price sensitive agents. Hence, when $\nu = -1$, information is taken up by the market faster than when $\nu =0$ and when $\nu = 1$. We also note that the phase ratio appears to exhibit saturation in the positive price sensitive agents, which should be expected.

\section{Generalization and Conjecture}
\begin{figure}[htbp]
\centering
\includegraphics[width=0.45\textwidth]{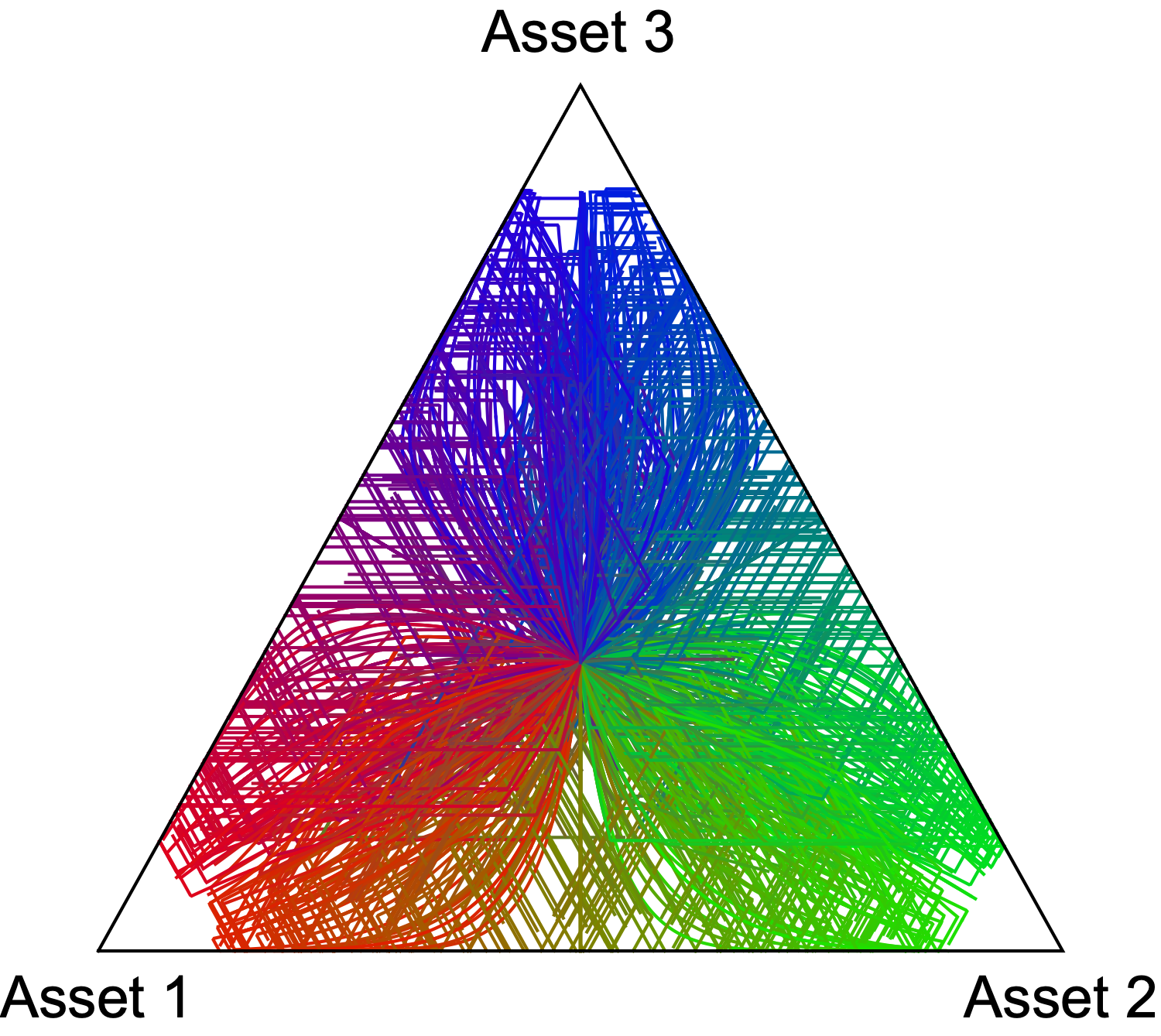}\qquad
\includegraphics[width=0.45\textwidth]{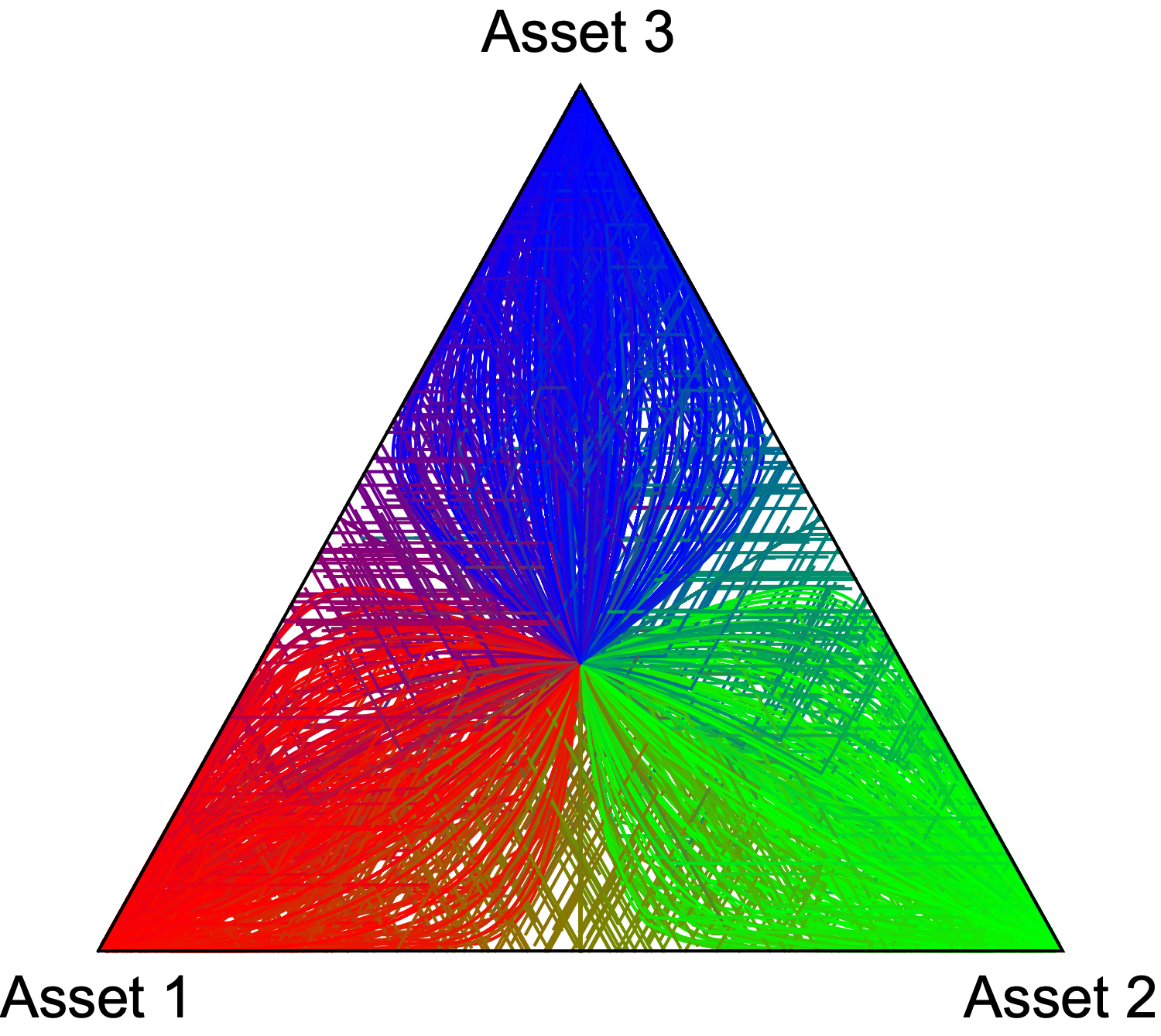}
\caption{(Left) $L_k$ randomly drawn from $[-2,2]$. (Right) $L_k$ randomly drawn from $[-5,5]$. In both cases the dynamics converge or asymptotically approach the extreme point of the unit simplex. No oscillations are observed.}
\label{fig:3Asset}
\end{figure}

Generalizing the dynamics of \cref{eqn:ODE} to $M>2$ assets is straightforward. We obtain:
\begin{equation}
\dot{p}_i =  \sum_{j=1}^M \beta p_i p_j\left(\dot{q}_i - \dot{q}_j\right),
\label{eqn:ODEN}
\end{equation}
where:
\begin{equation*}
\dot{q}_j = \sum_{k \in C_j} H\left\{\sigma\left[\psi_k(\mathbf{x},\mathbf{p})\right] - p_j\right\}.
\end{equation*}
Here $C_j$ is the group of agents who specialize in purchasing assets in class $j$ and $p_j$ is the price of asset class $j$. This equation is consistent with \cite{TVL03,BCFG22}. The dynamics effectively describe a recurrent neural network with an embedded softmax layer (generalizing \cref{fig:NN}). We conjecture that with constant information, the resulting dynamics always converge, fully generalizing the result in \cref{sec:Asymptotic}. We illustrate this conjecture in \cref{fig:3Asset} for a three asset market simulated with over 3000 variations on the dynamics assuming:
\begin{equation*}
\psi_k(\mathbf{x},\mathbf{p}) = L_k,
\end{equation*}
where $L_k$ is randomly drawn from either $[-2,2]$ or $[-5,5]$ - see \cref{fig:3Asset}(left) or \cref{fig:3Asset}(right), respectively. The cardinality $|C_j|$ is randomly drawn from $1$ to $10$, simulating a small market. 
The figures show that oscillation is not present when information is held constant. In particular, our conjecture implies that the dynamics quickly reach an $k < M$ dimensional Filippov surface and then slide along it until they reach a fixed point (a zero dimensional subset of the Filippov surface) thus ensuring convergence of the system. Assuming this conjecture is correct, then synthetic prediction markets (of the type discussed in \cite{NMGR21}) can be extended to arbitrary prediction classes and the use of continuous time markets will eliminate oscillations observed in the discrete time algorithm. In addition to the open question of the convergence of $M$-array option markets with $M>2$, we leave as an area of future research how best to design functions $\sigma$ and $\psi_i$ ($i=1,\dots,N$) to recover desired market output values given input data $\{\mathbf{x}_j\}_j$.

\section{Conclusion and Future Directions}\label{sec:Conclusions}
In this paper, we have prove the convergence of price in a continuous binary option market with constant external information. This is proved by showing that the right-hand-side of the dynamics are piecewise smooth and bounded allowing us to reason about the Filippov surface on which the dynamics come to rest. Consequently, the price dynamics are always monotonic, continuous, and piecewise smooth and always converge. In the presence of non-constant external information, we show empirically that a positive price sensitivity (i.e., purchases increase in response to higher prices) causes larger market lags to external information change while negative price sensitivity (i.e., purchases decrease  in response to higher prices) decrease market lag to external information. 

Results discussed in this paper can be extended in several ways. We have left an open conjecture on the generalization of the market convergence result to markets with $M>2$ asset classes. We assert that under constant external information these markets also converge. If this is true, then the use of continuous time $M$-array asset markets as an approach to machine learning would be sensible. In this case, since we have  shown $M$-array option markets describe a recurrent neural network, investigating how to fit exogenous parameters defining the functions $\varphi_i$ to build a classifier is an interesting and open question, especially considering the recurrent aspect of the neural structure. Finally further analysis in the case when the exogenous signal $\mathbf{x}(t)$ is non-constant may provide further insight into the behavior of these simple options markets. 

\section*{Acknowledgement}
Portions of this work were sponsored by the DARPA SCORE Program under Cooperative Agreement W911NF-19-2-0272.

\appendix
\section{The Lorenz Attractor} 
For the example in \cref{sec:Numerical}, we use the standard Lorenz attractor:
\begin{align*}
&\dot{x} = s\cdot\sigma(x-y)\\
&\dot{y} = s\cdot x\left(\rho - z - y\right)\\
&\dot{z} = s\cdot\left(xy - \alpha z\right).
\end{align*}
To better illustrate the market effect, we slow the dynamics down by a factor of $s = 0.05$ and set $\alpha = 1$, $\sigma = -3$, $\rho = 26.5$.
 
\section{Numerical Solution of Large Differential Equations}\label{sec:NumericalSoln}
When $\mathbf{x}(t)$ is complex or there are more than a few agents, numerical solution of the differential equations can become unstable. To compensate for this, we can solve \cref{eqn:ode1,eqn:ode0} using a simple Euler forward step method using a small $\epsilon$ in \cref{eqn:EulerStep} (see below):
\begin{displaymath}
q^1(t+\epsilon) \approx q^1(t) + \epsilon\sum_{i \in C_1} H\left\{\sigma[\psi_i(\mathbf{x},p)] - p(t)]\right\}.
\end{displaymath}
We then compute $p(t)$ using information for $q_1(t)$ and $q_0(t)$ up to time $t$ via \cref{eqn:dotp1}. For small $\epsilon \approx 0.01$, the results match the output of an off-the-shelf ODE solver and are more numerically stable.

\bibliographystyle{apsrev4-1}
\bibliography{main-ArXiv}
\end{document}